\documentclass[twoside,twocolumn,9pt]{article}
\pdfoutput=1
\usepackage{extsizes}
\usepackage[super,sort&compress,comma]{natbib} 
\usepackage[version=3]{mhchem}
\usepackage[left=1.5cm, right=1.5cm, top=1.785cm, bottom=2.0cm]{geometry}
\usepackage{balance}
\usepackage{times,mathptmx}
\usepackage{sectsty}
\usepackage{graphicx} 
\usepackage{lastpage}
\usepackage[format=plain,justification=justified,singlelinecheck=false,font={stretch=1.125,small,sf},labelfont=bf,labelsep=space]{caption}
\usepackage{float}
\usepackage{fancyhdr}
\usepackage{fnpos}
\usepackage[english]{babel}
\usepackage{array}
\usepackage[T1]{fontenc}
\usepackage[usenames,dvipsnames]{xcolor}
\usepackage{setspace}
\usepackage[compact]{titlesec}

\usepackage{sistyle}

\title{Capillary force on a tilted cylinder: AFM measurements} 

\author{Sebastien Kosgodagan Acharige\textit{$^{a}$}, Justine Laurent\textit{$^{a,b}$}, and Audrey Steinberger\textit{$^{a}$}$^{\ast}$}

\date{\today}

\begin{document}

\twocolumn[
\begin{@twocolumnfalse}

\maketitle


\begin{center}

$^{a}$~Univ Lyon, Ens de Lyon, Univ Claude Bernard Lyon 1, CNRS, Laboratoire de Physique, F-69342, Lyon, France
\\

$^{b}$~ESPCI, CNRS, PMMH, F-75005 Paris, France
\\

$^{\ast}$~E-mail: audrey.steinberger@ens-lyon.fr
\\
\end{center}

\vspace{2mm}

\begin{abstract}
We investigate the capillary force that applies on a tilted cylinder as a function of its dipping angle $i$, using a home-built tilting Atomic Force Microscope (AFM) with custom made probes. A micrometric-size rod is glued at the end of an AFM cantilever of known stiffness, whose deflection is measured when the cylindrical probe is dipped in and retracted from reference liquids. We show that a torque correction is necessary to understand the measured deflection. The results are compatible with a vertical capillary force varying as $1/\cos i$, in agreement with a recent theoretical prediction. We also discuss the accuracy of the method for measuring the surface tension times the cosine of the contact angle of the liquid on the hanging fiber.
\end{abstract}

\vspace{7mm}
\end{@twocolumnfalse}
]

\section{Introduction}

There are many situations where the capillary force or wetting properties matter when a slender object is interacting with a liquid interface. In the natural world for instance, water-walking insects can live atop the water surface thanks to the capillary force on their hydrophobic legs~\cite{hu2010}. On the contrary, this force has to be overcome by filamentous fungi living in moist environment, in order to grow aerial parts called hyphae that are necessary for the dissemination of their spores~\cite{Wosten}. On the applications side, wetting properties are of great importance for fiber and wire cleaning or coating processes~\cite{Quere_1999}, and for optimizing fog-harvesting nets~\cite{Park_fogharvesting}, while capillary interactions are responsible for self-assembly and collapse in hair assemblies or in microstructures fabrication~\cite{Roman_elastocap}, whether desirable or not.

Studying the capillary force exerted on a thin cylinder when it is immersed in and retracted from a fluid is also used as a measurement technique to characterize the surface tension $\gamma$ of the fluid and the wetting properties of the probe. This so-called micro-Wilhelmy technique, in which the cylindrical probes have a diameter well below the liquid's capillary length, has been initially developed using a microbalance to perform force measurements on vertical fibers having a diameter $D_0$ in the $20$ to $\SI{200}{\micro m}$ range. In this geometry, the capillary force $F_{cap}$ is simply equal to $\pi D_0 \gamma\cos\theta$, where $\theta$ is the contact angle of the fluid on the cylinder. This technique allowed to investigate the contact angle hysteresis due to the pinning and depinning of the contact line~\cite{di1990contact}, and also to measure the surface tension of highly viscous polymer melts~\cite{SAUER1991527}. More recently, it has been transposed to Atomic Force Microscope (AFM) measurements, where the cylindrical probe is attached to the end of an AFM cantilever~\cite{McGuiggan}. The very high force sensitivity of AFM cantilevers allows to perform capillary force and contact angle measurements on even smaller probes, with diameters ranging from a few micrometers down to about $\SI{10}{nm}$~\cite{Wang_PRE,yazdanpanah2008micro,Barber,delmas_PhysRevLett}. It also allows to investigate the dynamical properties of liquid meniscii at thermal equilibrium and under a large variety of solicitations~\cite{Jai_EPL2008,Tong_PRL2013,GuanPhysRevLett}. Another advantage of the AFM measurement scheme is the precise in-plane imaging and positioning that enables to probe very small amounts of fluids confined in tiny droplets or holes~\cite{Morris,Dupre}. However in the AFM experiments, the cylindrical probe is not always normal to the liquid interface. One may therefore wonder whether the orientation of the probe affects the AFM measurements, and what is the expected capillary force on a tilted cylinder.

Simple situations where an object is piercing a liquid perpendicularly to its surface are well understood~\cite{book}. However, situations were the liquid meniscus around the object is asymmetric, although very common, have been much less investigated. The 2D case of an elastic plate plunging with an inclination angle $i$ into a liquid bath has been studied in detail by Andreotti \emph{et al.}, showing a $1/\cos i$ dependence of the capillary force~\cite{Andreotti}. The case of a tilted cylinder has been numerically studied by Raufaste \emph{et al.}, first in a totally wetting situation~\cite{raufaste:hal-00725930} then with an arbitrary contact angle~\cite{Raufaste2013126}. In both cases, they have computed the meniscus shape and the resulting capillary force, and also retrieve a $1/\cos i$ dependence of the capillary force. They have experimentally checked the meniscus shape in the total wetting situation using a liquid jet piercing a soap film~\cite{raufaste:hal-00725930}, but up to our knowledge the capillary force on a tilted solid cylinder has never been checked so far.

In this work, we investigate experimentally the capillary force on a tilted cylinder using cylindrical AFM probes on a home-built tilting setup. We first describe this setup and the experimental procedure, then explain how the measured deflection can be properly related to the applied force, depending on the orientation of the probe. Finally, we compare our experimental results with the predictions of Raufaste \emph{et al.}~\cite{Raufaste2013126}, and discuss the accuracy of $\gamma\cos\theta$ measurements from these results.\\

\section{Materials and methods}

\begin{figure}[h]
	\centering
	\includegraphics[width=.35\textwidth]{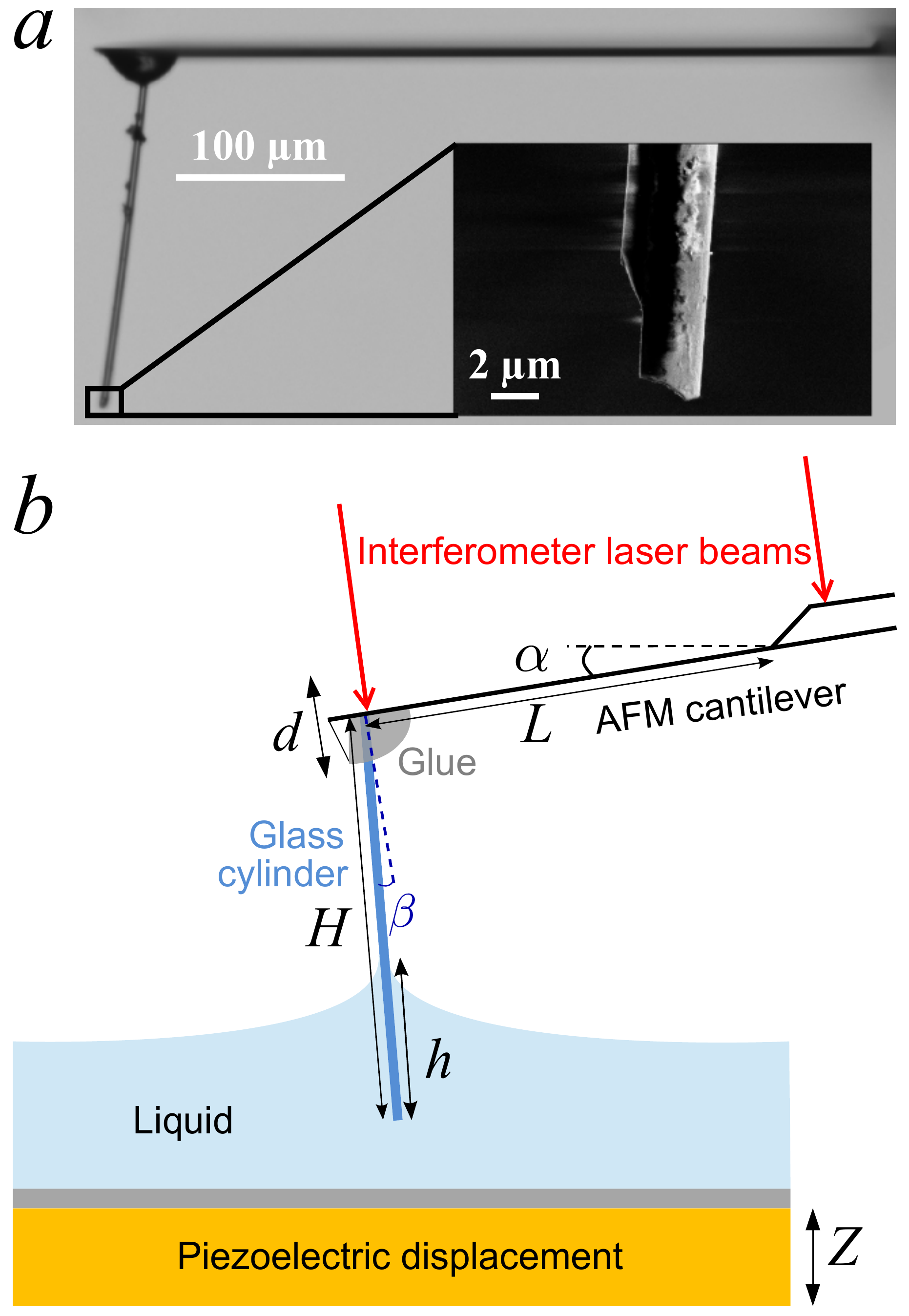}
	\caption{$ a $: Optical microscopy image (10x) of probe $ \# $ 3 in Table \ref{tbl:probes} viewed by the side. Insert: SEM image of the fiber's free end for the same probe. $ b $: Sketch of the probe immersed in a liquid.}
	\label{fgr:photolevier}
\end{figure}

All experiments were performed using a new home-built AFM with custom made probes. One of these probes can be seen in Figure~\ref{fgr:photolevier}$a$.
They are obtained using a technique already described elsewhere~\cite{devailly2014mode,de2016afm}.
A glass optical fiber is put under tension with a weight, then heated with a blowtorch to stretch it, and finally glued to the tip of a commercial AFM cantilever (NanoWorld Pointprobe ZEILR or Arrow FMR).
Three probes have been used in this work and their characteristics are listed in Table~\ref{tbl:probes}. The cantilever's length $ L $ between its clamped base and the apex of the fiber, the fiber's length $H$ and the angle $\beta$ between the perpendicular to the cantilever and the fiber over the last $\SI{100}{\micro m} $ at its free end, are measured from a side view using an optical microscope with a 10x magnification. The fiber's diameter at the free end $ D_0=D(h=0) $ and taper ratio $ c=\frac{\Delta D}{\Delta h} $ are measured using a Scanning Electron Microscope (SEM, Zeiss Supra 55VP) from a front view. The transverse angle between the normal to the cantilever and the fiber, as seen from the front view, has been checked to be below $2^{\circ}$ for all three probes. The spring constant $k$ of the cantilever is measured using the well-known thermal noise method~\cite{Butt_thermalnoise, Laurent}, from the fundamental resonance peak in the thermal noise spectrum of the deflection $d$ of the free cantilever.

\begin{table*}
	\small
	\caption{\ Characteristics of the AFM probes and corresponding symbols used in figures~\ref{fgr:y-intercept} and \ref{fgr:slope}}.
	\label{tbl:probes}
	\begin{tabular*}{\textwidth}{@{\extracolsep{\fill}}lllllll}
		\hline
		Probe $ \# $ and symbol & $k~\SI{}{(N.m^{-1})}$ & $ L~\SI{}{(\micro m)}$ & $H~\SI{}{(\micro m)}$& $\beta$ & $D_0~\SI{}{(\micro m)}$ & $ c $\\
		\hline
		1 - Triangles & $ 3.9\pm0.4 $ & $ 236\pm5 $ &$ 271\pm5 $&$ 7.3^{\circ}\pm1.0 $&$ 2.59\pm0.05 $&$(+1.3 \pm 0.7)\times10^{-3}$\\
		2 - Circles & $ 1.5\pm0.2 $ & $ 444\pm5 $ &$ 185\pm5 $&$ -7.0^{\circ}\pm1.0 $&$ 2.70\pm0.05 $&$ (-2.0 \pm 0.7)\times10^{-3}$\\
		3 - Squares & $ 1.7\pm0.2 $ & $ 436\pm5 $ &$ 203\pm5 $&$ 7.0^{\circ}\pm1.0 $&$ 4.01\pm0.05 $&$ (-1.6 \pm 0.7)\times10^{-3}$\\
        \hline
	\end{tabular*}
\end{table*}

We developed an innovative AFM by building the whole head of the instrument (probe and detection) on a rotating frame. Thus, the angle $\alpha$ of the cantilever with the horizontal plane, as displayed on Figure \ref{fgr:photolevier}, can be controlled. While for an usual AFM, this angle $ \alpha $ is fixed and lies between $ 11^{\circ} $ and $ 15^{\circ} $, it can be set from $ 5^{\circ} $ to $ 30^{\circ} $ in our setup.

In this dedicated home-built AFM, the deflection $d$ at the apex of the fiber is directly measured using a quadrature-phase interferometric sensor which was designed by Paolino \emph{et al.} \cite{Paolino_interferometer}. This sensor measures with a very high accuracy the optical path difference between a reference laser beam reflecting of the static base of the cantilever and a sensing beam reflecting at the apex of the fiber, as depicted in Figure~\ref{fgr:photolevier}. The main advantage here is that this deflection measurement is intrinsically calibrated with respect to the $\SI{632.8}{nm} $ laser wavelength, thus requiring no further calibration with the functionalized probe contrary to the optical lever technique that is used in most AFM setups.

Three reference liquids were used for the experiments, two alkanes (decane and dodecane) and a silicone oil, who were chosen for their low contact angle hysteresis on glass substrates. They all have a wetting angle $ \theta $ close or equal to $ 0 $ on glass, and slightly different surface tensions $\gamma$. Their physical properties are listed in Table \ref{tbl:liquids}.
For decane and dodecane, the value of $ \gamma\cos\theta $ was measured through capillary rise in glass capillaries.

The liquid fills a small glass container having a $\SI{2}{mm} $ height and a $\SI{18}{mm} $ inner diameter, placed under the cylindrical probe on a piezoelectric stage. When changing the liquid, the container is cleaned with soap and water then sonicated in isopropyl alcohol and finally rinsed first with deionized water and then with the liquid of interest. The silicone oil is always used last. The measurements are performed at $ 21.5^{\circ}\mathrm{C} $ in an air-conditioned room.

\begin{table}[h]
	\small
	\caption{\ Tabulated surface tension $\gamma$, density $\rho$ and viscosity $\eta$ of the three used liquids, and value of $\gamma\cos\theta$ (with $\theta$ the wetting angle) on glass for decane and dodecane as measured by capillary rise experiments. All experiments took place at $ 21.5^{\circ}\mathrm{C} $. }
	\label{tbl:liquids}
	\begin{tabular*}{0.48\textwidth}{@{\extracolsep{\fill}}lllll}
		\hline
		Liquid  & $ \gamma~\SI{}{(mN.m^{-1})} $ & $ \gamma\cos\theta~\SI{}{(mN.m^{-1})} $ & $\rho~\SI{}{(kg.m^{-3})} $ & $\eta~\SI{}{(mPa.s)}$ \\
		\hline
		Decane & $ 23.6 $ & $ 23.4 \pm 1.2$ & $ 730 $ & $ 0.9 $ \\
		Dodecane & $ 25.2 $ & $ 24.3 \pm 1.2$ & $ 750 $ & $ 1.3 $ \\
		Silicone oil & $ 20.7 $ & $ 20.7 $ & $ 960 $ & $ 50 $\\
		\hline
	\end{tabular*}
\end{table}

The experiments take place as follows: the piezoelectric stage performs an approach and retract over a $\SI{100}{\micro m} $ vertical range at a constant speed: $\SI{100}{\micro m.s^{-1}} $ for the alkanes and $\SI{10}{\micro m.s^{-1}} $ for the silicone oil. This speed is chosen so that the viscous drag on the fiber remains small for all the investigated liquids, while the change of liquid height due to evaporation is negligible over the time of an approach and retract (less than $\SI{100}{nm} $ over $\SI{2}{s} $ for the most volatile liquid, decane).
During this operation, the cantilever deflection $ d $ and height of the piezoelectric stage $ z $ are recorded simultaneously
using a NI PXIe-6368 acquisition card.
This process is repeated ten times for a given angle $ \alpha $, which is then changed to perform a new series of ten acquisitions.
For a given probe and liquid, the angle was varied on the full possible range, from $ 5^{\circ} $ to $ 30^{\circ} $ by $ 5^{\circ} $ steps.

\begin{figure}[h]
	\centering
	\includegraphics[width=.45\textwidth]{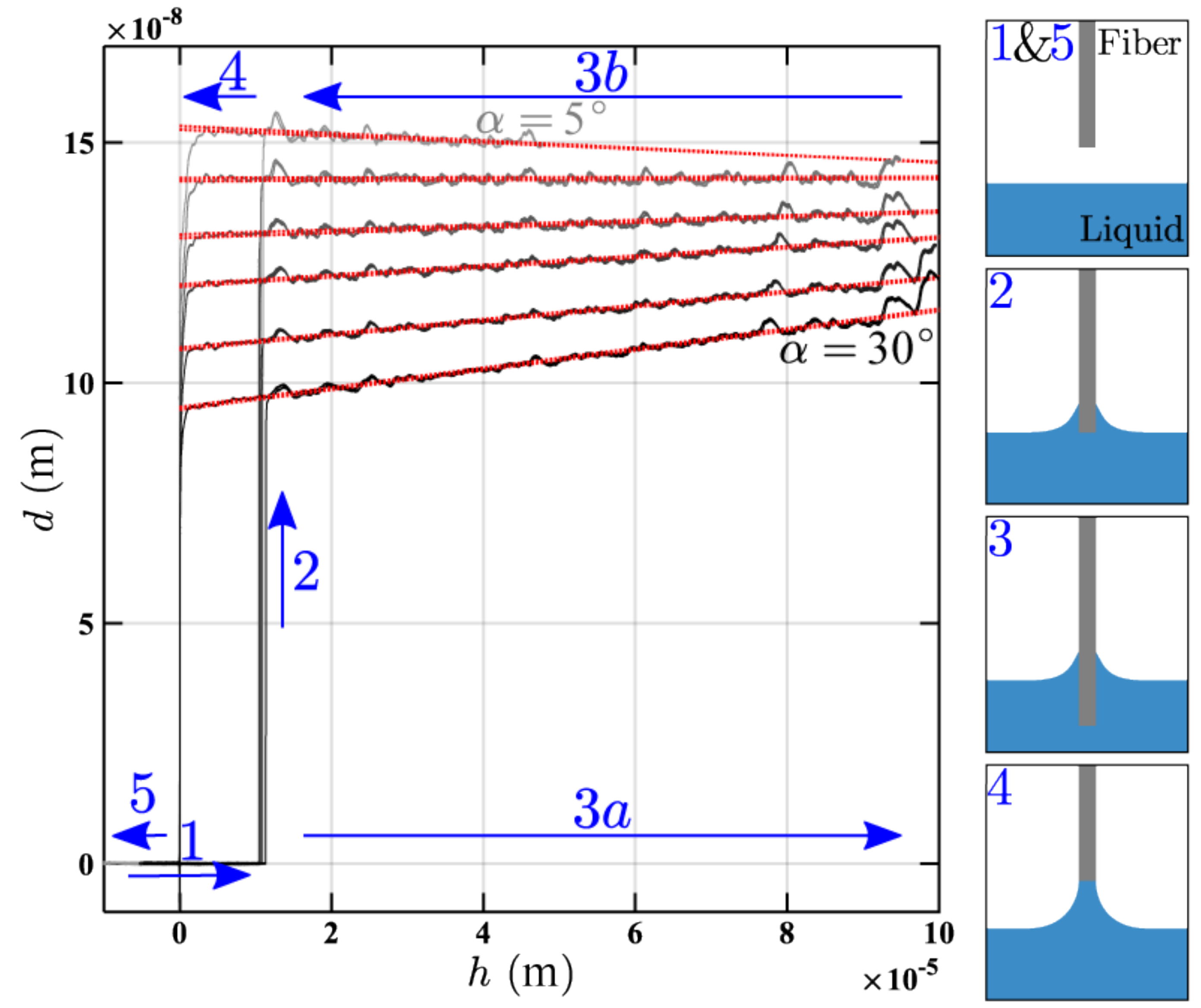}
	\caption{Deflection $d$ as a function of immersed fiber length $ h $, for various angles $ \alpha $. The data was obtained with probe $ \#\ 3$ in Table \ref{tbl:probes} in silicone oil. Red dotted lines are linear fits of the displacement for the phases $ 3a $ and $ 3b $. On the right side are sketches of the fiber position during the experiment, with corresponding numbers in the graph.}
	\label{fgr:FCs}
\end{figure}

Following the above protocol gives a set of displacement curves, averaged over ten acquisitions, plotted as a function of the immersion length $ h=\frac{z}{\cos i} $, as shown in Figure \ref{fgr:FCs}. 
During the approach, the deflection $d$ remains equal to zero as long as the fiber doesn't touch the liquid (phase 1). When the fiber enters in contact with the liquid, a meniscus forms and applies a capillary force on the probe (phase 2). When dipping the fiber further, the deflection $d$ varies linearly with the dipping depth $h$, except for small reproducible deviations which are due to imperfections on the fiber's surface averaged over the perimeter of the contact line (phase 3). These deviations get smaller as the immersion angle $i$ increases because of the increase of the averaging perimeter. Upon retraction, one can observe a hysteresis while the meniscus slides beyond the position where it has first formed upon phase 2, with sharper deviations due to geometrical features arising the fiber's cutting procedure (phase 4), until breakup when the meniscus detaches from the end of the fiber and the capillary force disappears (phase 5). For all curves, the origin of the distance $ h=0 $ is set at the meniscus breakup (separating phase 4 and 5). 

During phase 3, the experimental curves can be fitted with a linear equation to recover the slope $ s=\frac{\partial d}{\partial h} $ and y-intercept $ d_0=d(h=0) $ for each angle $\alpha$. The approach phase  $ 3a $ and retract phase $ 3b $ are fitted separately. The change of slope $s$ between approach and retraction remains very small (about $5\%$), confirming that the viscous drag is small as expected given the chosen vertical speed. The y-intercept $ d_0 $ are also very close between approach and retraction, showing that the contact angle hysteresis is small, which is also expected given the chosen liquids and vertical speed.
However, the slope $s$ and the y-intercept $d_0$ depend strongly one the tilting angle $\alpha$ for a given probe and a given liquid, as can be clearly seen on Figure \ref{fgr:FCs}. To understand these features, the mechanics of the probes is considered in next section.\\

\section{Theory}
\label{sec_theory}

\begin{figure}[h]
	\centering
	\includegraphics[width=.35\textwidth]{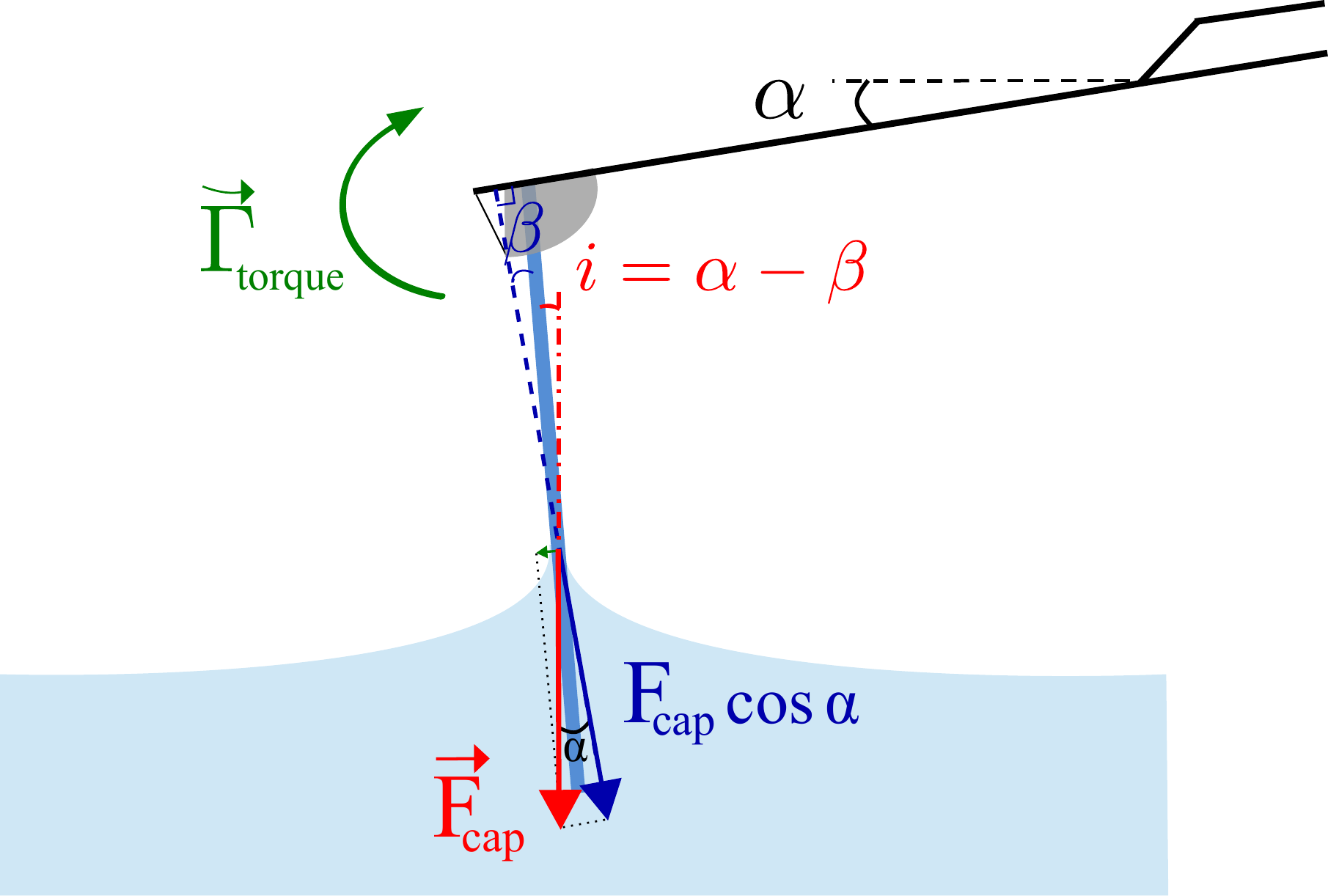}
	\caption{Sketch of the forces and torque applied to the system.}
	\label{fgr:meca}
\end{figure}

Let's consider a probe whose fiber is dipped in a liquid with an immersion angle $i=\alpha -\beta$, as sketched in Figure \ref{fgr:meca}.
The liquid meniscus applies a capillary force $ \vec{F}_{cap} $ on the fiber, at a distance $H-h$ from its fixing point on the cantilever~\cite{buoyancy}.
The total deflection $d$ of the cantilever measured by the interferometric sensor is the sum of two components. The first, denoted $d_{force}$, results from the pulling force due to the component of the capillary force that is perpendicular to the cantilever. The second, denoted $d_{torque}$, results from the torque exerted on the cantilever by the component of the capillary force that is perpendicular to the fiber.

Now, what is the direction of the capillary force on a tilted cylinder, where the meniscus doesn't display a rotational symmetry? 
In their numerical work~\cite{Raufaste2013126}, Raufaste and Cox show that the capillary force remains vertical for a meniscus at equilibrium in an infinite bath. 
This can be understood using a simple energetic argument. When the meniscus is at equilibrium, there is no surface energy change upon a horizontal translation of the fiber in an infinite bath, thus no work of the capillary force, and hence the capillary force has no horizontal component. 
With a vertical capillary force, the total deflection arising from the pulling force $F_{cap}\cos\alpha$ and the torque $ \Gamma_{torque}= (H-h)F_{cap}\sin i $ is~\cite{Edwards}:
\begin{equation}
\label{d_h}
d(h)=d_{force}+d_{torque}=\frac{F_{cap}}{k}\left(\cos\alpha-\dfrac{3}{2}\dfrac{H-h}{L}\sin i\right)
\end{equation}

Since $F_{cap}$ doesn't depend on $h$ on a cylinder in absence of contact line pinning, equation~\ref{d_h} shows that $ d(h) $ is a linear function of $ h $, which corresponds indeed to the behaviour observed in Figure \ref{fgr:FCs}. It also depends on the tilting angle through the $\cos\alpha$ and $\sin i$ terms and the capillary force $F_{cap}$, which also depends on the immersion angle $i$. According to Raufaste and Cox~\cite{Raufaste2013126},
\begin{equation}
\label{F_cap}
F_{cap}=\dfrac{\pi D_0 \gamma\cos\theta}{\cos i}
\end{equation}

Plugging this expression into equation~\ref{d_h} and noting that $\alpha=i+\beta$, yields the following expressions for the y-intercept $d_0(i)$ and the slope $s(i)$:
\begin{equation}
\label{d_0_cyl}
d_0(i)=\dfrac{\pi D_0 \gamma\cos\theta}{k}\left(\cos\beta-\tan i\left(\sin\beta+\frac{3H}{2L}\right)\right)
\end{equation}

\begin{equation}
\label{s_cyl}
s(i)\!=\!\dfrac{3\pi D_0 \gamma\cos\theta}{2Lk}\!\tan i 
\end{equation}

In practice, the fibers are not perfectly cylindrical but slightly conical, as confirmed by the SEM pictures and characterized by the taper ratio $ c $ in Table \ref{tbl:probes}. Since the taper ratio $c$ is small, we model this by replacing the diameter $ D_0 $ by the local diameter $ D(h)=D_0(1+\frac{hc}{D_0}) $ in equation~\ref{F_cap}, before plugging it in equation~\ref{d_h}. Neglecting the second order term, the expression for the slope $s(i)$ becomes:
\begin{equation}
\label{s_conical}
s(i)\!\simeq\!\dfrac{3\pi D_0 \gamma\cos\theta}{2Lk}\!\left(\!\tan i \left(\!1\!-\!\frac{2Lc}{3D_0}\!\left(\!\sin\beta\!+\!\frac{3H}{2L}\right)\!\!\right)\!+\!\frac{2Lc}{3D_0}\cos\beta\!\right),
\end{equation}
while the expression for $d_0(i)$ is not modified. Both $d_0(i)$ and $s(i)$ are thus expected to be linear with $\tan i$ with a non zero y-intercept if equation~\ref{F_cap} is true. Equations \ref{d_0_cyl} and \ref{s_conical} may then be used to measure $\gamma\cos\theta$ if the spring constant $k$ and the geometrical parameters of the probe are known. Note that equation~\ref{d_0_cyl} reduces to the usual expression for the Wilhelmy technique in the limit where $i=0$.\\

\section{Results and discussion}
\label{sec_discussion}

In the following, the slopes $s$ obtained at approach and retraction for a given set of parameters are averaged together, in order to remove any residual hydrodynamical effect. This is justified by the low Reynolds number of these experiments, as the highest attained value is $ 3.2\times10^{-4} $ for probe $ \# 3 $ and decane. The y-intercepts $ d_0 $ at approach and retraction are also averaged, meaning that the dynamic contact angle hysteresis is neglected in the following data analysis. This is justified because it remains comparable with the uncertainty on the $ d_0 $ values obtained from the linear fit of the deflection $d$ versus immersed length $h$ data, as can be seen on figure~\ref{fgr:FCs}.

\begin{figure}[h]
	\centering
	\includegraphics[width=.45\textwidth]{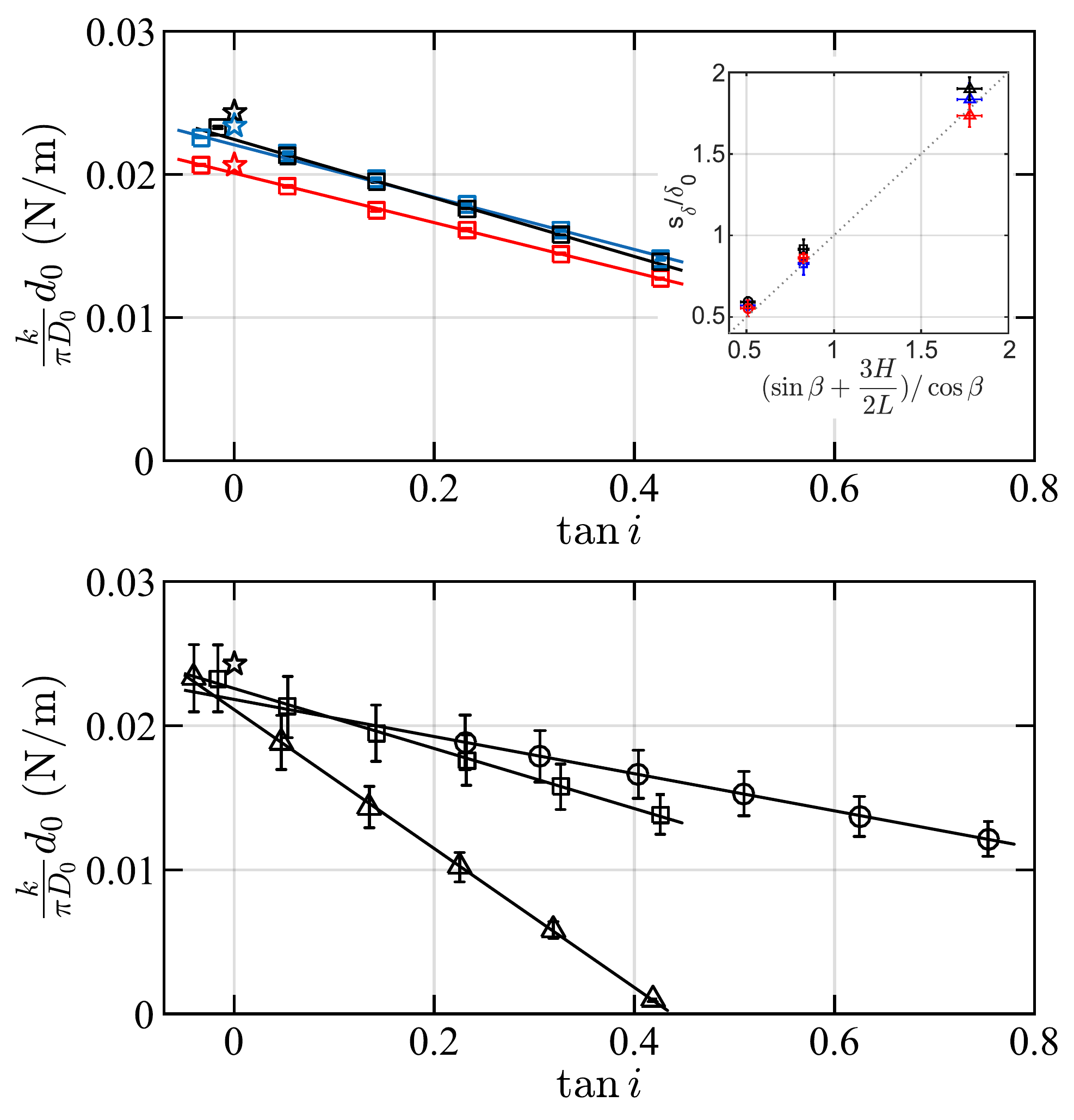}
	\caption{Normalized y-intercept $ \frac{kd_0}{\pi D_0}$ as a function of $ \tan i $, for decane (blue), dodecane (black) and silicone oil (red) on probe $ \#\ 3$ (top), and for dodecane on the 3 probes (bottom). Probe symbols are given in table \ref{tbl:probes}. Solid lines are linear fits and stars are expected values for $ \gamma\cos\theta $. Inset: comparison of $s_{\delta}/\delta_0$ with $(\sin\beta + \frac{3H}{2L})/\cos\beta$ for all the data sets, with a dotted line of slope 1 as a guideline.}
	\label{fgr:y-intercept}
\end{figure}

\begin{figure}[h]
	\centering
	\includegraphics[width=.45\textwidth]{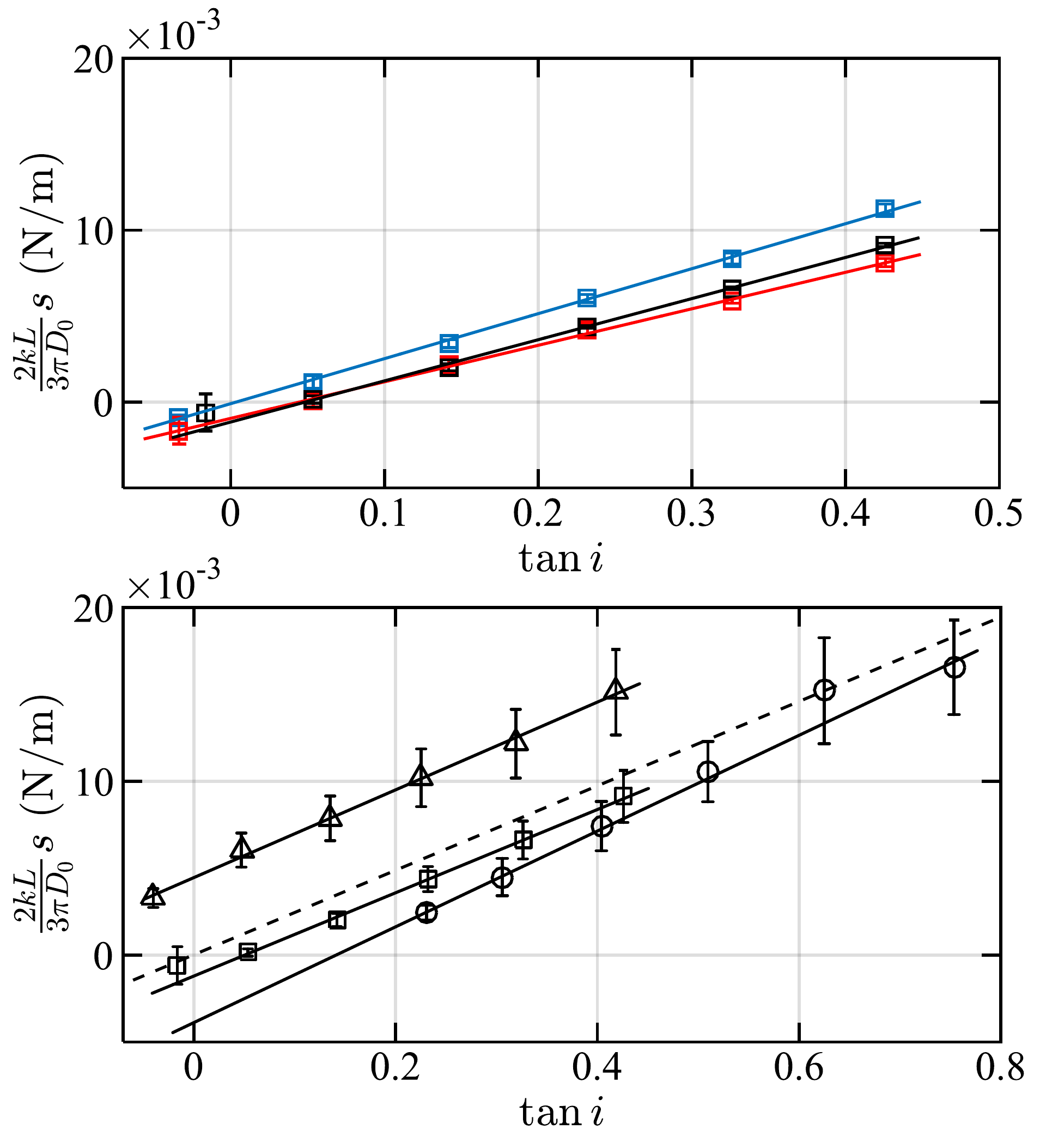}
	\caption{Normalized slope $ \frac{2Lks}{3\pi D_0}$ as a function of $ \tan i $, for decane (blue), dodecane (black) and silicone oil (red) on probe $ \#\ 3$ (top), and for dodecane on the 3 probes (bottom). Probe symbols are given in table \ref{tbl:probes}. Solid lines are linear fits. The dashed line has a slope equal to the expected $ \gamma\cos\theta $ value for dodecane.}
	\label{fgr:slope}
\end{figure}

The normalized y-intercept $ \frac{kd_0}{\pi D_0}$ is plotted as a function of $ \tan i $ in figure~\ref{fgr:y-intercept}, and the normalized slope $ \frac{2Lks}{3\pi D_0}$ is plotted as a function of $ \tan i $ in figure~\ref{fgr:slope}. On both figures, the top panel represents the data obtained for the three liquids with the same probe (probe $ \#\ 3$ in table~\ref{tbl:probes}). The displayed error bars correspond to the uncertainty resulting from the fitting procedure only, not taking into account the uncertainty on the normalization factor. The bottom panel represents the data obtained for the same liquid (dodecane), with the three probes. There, the displayed errors bars also include the uncertainty on the normalization factor. They are dominated by the uncertainty on the cantilever's spring constant $k$, which is about $10\%$ of the measured value. As can be seen on both figures, the data points for each experiment align well on straight lines in agreement with the behavior predicted by equations~\ref{d_0_cyl} and \ref{s_conical}, thus in agreement with a $\frac{1}{\cos i}$ behaviour of the capillary force $F_{cap}$ on a tilted cylinder with an immersion angle~$i$.

Let's now have a closer look on the $ \frac{kd_0}{\pi D_0}$ versus $ \tan i $ data plotted in figure~\ref{fgr:y-intercept}. They are fitted by the linear equation $ \frac{kd_0}{\pi D_0} = \delta_0 - s_{\delta}\tan i$.
One can first compare the $s_{\delta}/\delta_0$ ratio with $(\sin\beta + \frac{3H}{2L})/\cos\beta$, a parameter which depends only on the probes' geometry and not on the magnitude of $F_{cap}$. The satisfying agreement shown in the inset of figure~\ref{fgr:y-intercept} further confirms the mechanical model proposed in section~\ref{sec_theory} to analyze the probes' deflection.

The y-intercepts $\delta_0$ are close (within error bars) to the $\gamma\cos\theta$ values of the different liquids given in table~\ref{tbl:liquids} and represented as stars on the y-axis of figure~\ref{fgr:y-intercept}. This is in quantitative agreement with equation~\ref{d_0_cyl} that predicts $\delta_0 = \gamma\cos\theta\cos\beta$, since the $\cos\beta$ values are very close to 1 for our probes. The $\delta_0/\cos\beta$ ratios, denoted $ \gamma_1$, are given in table~\ref{tbl:gamma}. They compare very well with the $\gamma\cos\theta$ values in table~\ref{tbl:liquids}, showing that this ratio is a good way to measure $ \gamma\cos\theta $. The uncertainty indicated in table~\ref{tbl:gamma} is dominated by the $10\%$ uncertainty on the spring constant $k$ in the normalization factor. It is possible to get rid of the error on the probe's parameters by cross-multiplying the results with a reference fluid of known $\gamma\cos\theta$ on the same probe. Here, using silicone oil as the reference, the uncertainty falls down to about $4\%$, as shown in column $\gamma_{1\times}$ of table~\ref{tbl:gamma_x}. This remaining uncertainty comes mainly from the error bar resulting from the fitting procedures, which is of the order of $2\%$ for the $\delta_0$ parameter. A smaller contribution due to a repositioning uncertainty of the probe between two experiments, evaluated to $\pm 0.5^{\circ}$ on the zero of the immersion angle $i$, is also taken into account.

\begin{table}[h]
	\small
	\caption{\ Comparison of the 3 ways of measuring $\gamma\cos\theta$ from the experimental data: $ \gamma_1=\delta_0/\cos\beta$,  $\gamma_2=\frac{s_{\delta}}{\sin\beta +\frac{3H}{2L}}$ and $\gamma_3=\frac{s_{\Sigma}}{1-\frac{2Lc}{3D_0}\!\left(\!\sin\beta\!+\!\frac{3H}{2L}\right)\!}$. }
	\label{tbl:gamma}
	\begin{tabular*}{0.48\textwidth}{@{\extracolsep{\fill}}lllll}
		\hline
		Liquid  & Probe $ \# $ & $ \gamma_1~\SI{}{(mN.m^{-1})} $ & $ \gamma_2~\SI{}{(mN.m^{-1})} $& $ \gamma_3~\SI{}{(mN.m^{-1})}$ \\
		\hline
		       & 1 & $ 21.2 \pm 2.5$ & $ 24.7 \pm 3.0$ &$ 28.8 \pm 4.4$ \\
		Decane & 2 & $ 21.1 \pm 2.3$ &$ 23.7 \pm 3.3$ & $ 22.7 \pm 3.0$ \\
		       & 3 & $ 22.3 \pm 2.3$ & $ 22.3 \pm 3.1$ & $ 23.9 \pm 3.1$ \\
		\hline
	             & 1 & $ 21.6 \pm 2.4$ & $ 26.1 \pm 3.1$ & $ 27.5 \pm 4.4$ \\
		Dodecane & 2 & $ 22.0\pm 2.4$ & $ 25.6 \pm 3.3$ & $ 24.8 \pm 3.7$ \\
		         & 3 & $ 22.6 \pm 2.3$ & $ 25.0 \pm 3.3$ & $ 21.8 \pm 3.1$ \\
		\hline
                    & 1 & $ 19.4 \pm 2.1$ & $ 21.2 \pm 2.5$ & $ 22.8 \pm 3.2$\\
		Silicone oil & 2 & $ 18.0 \pm 2.2$ & $ 19.7\pm 3.2$ & $ 16.4 \pm 2.6$ \\
		            & 3 & $ 20.3 \pm 2.1$ & $ 21.1 \pm 2.5$ & $ 19.4 \pm 2.6$ \\
		\hline
	\end{tabular*}
\end{table}

\begin{table}[h]
	\small
	\caption{\ Comparison of the 3 ways of measuring $\gamma\cos\theta$ from the experimental data, using silicone oil as a reference fluid: $\gamma_{n\times}=20.7\gamma_n(\mathrm{alkane})/\gamma_n(\mathrm{silicone~oil})$. }
	\label{tbl:gamma_x}
	\begin{tabular*}{0.48\textwidth}{@{\extracolsep{\fill}}lllll}
		\hline
		Liquid  & Probe $ \# $ & $ \gamma_{1\times}~\SI{}{(mN.m^{-1})} $ & $ \gamma_{2\times}~\SI{}{(mN.m^{-1})} $& $ \gamma_{3\times}~\SI{}{(mN.m^{-1})}$ \\
		\hline
		       & 1 & $ 22.6 \pm 0.8$ & $ 24.1 \pm 1.4$ &$ 26.1 \pm 2.6$ \\
		Decane & 2 & $ 24.2 \pm 1.0$ &$ 24.9 \pm 2.5$ & $ 28.7 \pm 2.9$ \\
		       & 3 & $ 22.8 \pm 0.7$ & $ 21.9 \pm 1.9$ & $ 25.6 \pm 2.6$ \\
		\hline
	             & 1 & $ 23.0 \pm 0.8$ & $ 25.5 \pm 1.5$ & $ 25.0 \pm 2.7$ \\
		Dodecane & 2 & $ 25.2\pm 1.0$ & $ 27.0 \pm 2.5$ & $ 31.3 \pm 3.5$ \\
		         & 3 & $ 23.1 \pm 0.9$ & $ 24.5 \pm 1.7$ & $ 23.3 \pm 2.3$ \\
		\hline
	\end{tabular*}
\end{table}

According to equation~\ref{d_0_cyl}, another way to retrieve $ \gamma\cos\theta $ is to compute $\gamma_2=\frac{s_{\delta}}{\sin\beta +3H/2L}$, see table~\ref{tbl:gamma}. The results are again compatible with the expected values within the error bars. 
However, the obtained results tend to be slightly over-evaluated (by a few percents) due to the choice of setting the origin of the immersion depth $h=0$ at the meniscus breakup. 
This source of systematic error is not included in the uncertainty evaluated in table~\ref{tbl:gamma}, of about $13\%$, which is mainly due to the error on the probe's parameters. 
The error coming from the fitting procedures on the slope $s_{\delta}$ is of the order of $5\%$ on average, so that the error on $ \gamma\cos\theta $ only reduces to about $8\%$ when using the silicone oil values as a reference, as shown in table~\ref{tbl:gamma_x}, column  $\gamma_{2\times}$. As a result, this second way to retrieve $ \gamma\cos\theta $ is less accurate than the one discussed in the previous paragraph. 

On figure~\ref{fgr:slope}, the normalized slope $\frac{2Lks}{3\pi D_0}$ versus $ \tan i $ data are fitted by the linear equation $\frac{2Lks}{3\pi D_0} = \Sigma_0 + s_{\Sigma}\tan i$. The non-zero y-intercepts $\Sigma_0$ can be explained by the small but non-zero taper ratio $c$ of the fibers, which leads to $ \Sigma_0 = \gamma\cos\theta\cos\beta\frac{2Lc}{3D_0}$ according to equation~\ref{s_conical}. From the taper ratio values given in table~\ref{tbl:liquids}, one can check that the magnitude of $\Sigma_0$ is reasonable within this picture (of the order of a few mN/m), and that they have the expected signs and fall in the expected order with respect to each other on the bottom panel of figure~\ref{fgr:slope}. The very large uncertainty on the geometrical measurement of $c$ and the similar one on the $\Sigma_0$ value make it hard to perform a more quantitative comparison.

According to equation~\ref{s_conical}, the slope $s_{\Sigma}$ is very close to $ \gamma\cos\theta $, with only a small correction of the order of $10\%$ due to the small taper ratio of the fiber. On the bottom panel of figure~\ref{fgr:slope}, the slopes of the experimental data compare indeed well with the dashed line of slope $ \gamma\cos\theta $ for dodecane. A third way to retrieve $ \gamma\cos\theta $ from the deflection data is thus to compute $\gamma_3=\frac{s_{\Sigma}}{1-\frac{2Lc}{3D_0}\!\left(\!\sin\beta\!+\!\frac{3H}{2L}\right)\!}$. The results are given in table~\ref{tbl:gamma} and are again compatible with the expected values within the error bars. The relative uncertainties are quite similar to the one associated with the second method using the slope $s_{\delta}$. It barely reduces to about $10\%$ when using the silicone oil values as a reference (see $\gamma_{3\times}$ in table~\ref{tbl:gamma_x}, so that there is no real gain in using a reference fluid with this third method.

By comparing the different ways to measure $ \gamma\cos\theta $ from the data sets in tables~\ref{tbl:gamma} and \ref{tbl:gamma_x}, one can see that the most accurate one is by looking at $ \gamma_{1\times}$, that's to say the first method with silicone oil as a reference fluid. The measurement precision can of course be further increased by averaging the data over the three fibers, yielding $ \gamma\cos\theta = (23.3 \pm 0.6)~\SI{}{mN.m^{-1}} $ for decane and $ \gamma\cos\theta = (24.5 \pm 0.6)~\SI{}{mN.m^{-1}} $ for dodecane on our glass fibers.\\

\section{Conclusion}

In this work, we use a specifically built tilting AFM with custom probes to measure the capillary force exerted on a solid cylinder as a function of its immersion angle $i$. 
We show that, due to the large $H/L$ ratio of the hanging-fiber probes, the torque applied to the cantilever by the capillary force is not negligible, and has to be taken into account to understand quantitatively the measured deflection. It is therefore necessary to evaluate this torque contribution when using AFM-based techniques to study wetting properties such as surface tension, contact angle or contact line pinning. This contribution can be safely neglected when $\frac{H}{L}\sin i \ll 1$, for example when using nanometric probes with a $H/L \ll 1$ ratio~\cite{Dupre} or micrometric but vertical fibers~\cite{GuanPhysRevLett,Wang_PRE}.

With this torque correction, the experimental data agree nicely with Raufaste and Cox's prediction for the capillary force applied by a meniscus at equilibrium on a tilted cylinder~\cite{Raufaste2013126}, that's to say a vertical capillary force equal to $\frac{\pi D_0 \gamma\cos\theta}{\cos i}$. As a result, this good agreement validates their numerical approach.

In these experiments, the very small but non vanishing taper ratio of the probes could be detected when analysing the slopes $s$ as a function of the tilting angle $i$.
Our tilting setup could therefore be further used to study the capillary force on tilted cones or pyramids,
which is of interest for drop collection on conical structures\cite{ju,Lorenceau} and for many AFM-based investigations of liquid meniscii. Indeed, most of AFM tips have a conical or a pyramidal shape and are slanted with respect to the interface, leading to asymmetric nano-meniscii on tilted cones or pyramids~\cite{Chau,Koeber,Jai_EPL2008,delmas_PhysRevLett}.

Finally, the hanging-fiber AFM technique yields several ways to measure $ \gamma\cos\theta $ from the deflection data. The most accurate one is to measure the y-intercept of a linear fit of the deflection data as a function of the dipping depth $h$, extrapolate it to $i=0$, and normalize the resulting value using a reference fluid such as silicone oil to eliminate the uncertainties on the probe's parameters. This technique could be used to measure the wetting properties of different kinds of hairs and fibers. Even more interestingly, thank to the very high sensitivity of AFM deflection measurements, it could also be useful to measure locally weak surface tension changes and/or weak surface tension at liquid/liquid interfaces.\\

\section*{Acknowledgements}

We thank E. Charlaix, B. Andreotti, and B. Pottier for fruitful discussions and R. Taub for her help with the setup. We acknowledge the financial support of the ANR project NANOFLUIDYN (Grant No. ANR-13-BS10-0009) and the ERC project OutEFLUCOP.\\

\renewcommand\refname{Notes and references}

\bibliography{capForce_Kosgodagan_arXiv} 

\begin{thebibliography}{10}
\bibitem{buoyancy}
{The Archimedes buoyancy force on the immersed part of the fiber is completely
  negligible here, as it is about 5 orders of magnitude smaller than the
  capillary force on these micrometric fibers.}
\end{thebibliography}
\bibliographystyle{ieeetr}

\end{document}